\documentclass[aps,prl,twocolumn,superscriptaddress,showpacs]{revtex4}

\pdfoutput=1

\def\({\left(}
\def\){\right)}
\def\[{\left[}
\def\]{\right]}
\usepackage{amsmath}
\usepackage{amsfonts}
\usepackage{amssymb}
\usepackage{graphicx}
\usepackage{epsfig}
\usepackage{subfig}
\begin{document}


\title{Spinodal Instabilities and Super-Planckian Excursions in Natural Inflation}


\author{Andreas Albrecht}
\email{ajalbrecht@ucdavis.edu}
\affiliation{University of California at Davis, Department of Physics, One Shields Ave, Davis CA 95616 USA}
\author{R.~Holman}
\email{rh4a@andrew.cmu.edu}
\affiliation{Physics Department, Carnegie Mellon University, Pittsburgh PA 15213, USA}
\author{Benoit J. Richard}
\email{bjrichard@ucdavis.edu}
\affiliation{University of California at Davis, Department of Physics, One Shields Ave, Davis CA 95616 USA}


\date{\today}

\begin{abstract}
Models such as Natural Inflation that use Pseudo-Nambu-Goldstone bosons (PNGB's) as the inflaton are attractive for many reasons. However, they typically require trans-Planckian field excursions $\Delta \Phi>M_{\rm Pl}$, due to the need for an axion decay constant $f>M_{\rm Pl}$ to have both a sufficient number of e-folds {\em and} values of $n_s,\ r$ consistent with data.  Such excursions would in general require the addition of all other higher dimension operators consistent with symmetries, thus disrupting the required flatness of the potential and rendering the theory non-predictive. We show that in the case of Natural Inflation, the existence of spinodal instabilities (modes with tachyonic masses) can modify the inflaton equations of motion to the point that versions of the model with $f<M_{\rm Pl}$ can still inflate for the required number of e-folds. The instabilities naturally give rise to two separate phases of inflation with different values of the Hubble parameter $H$, one driven by the zero mode, the other by the unstable fluctuation modes. The values of $n_s$ and $r$ typically depend on the initial conditions for the zero mode, and, at least for those examined here, the values of $r$ tend to be unobservably small. 
\end{abstract}

\pacs{98.80.Es, 98.80.Cq}

\maketitle

While the inflationary paradigm is consistent with all data coming from the CMB\cite{Ade:2015lrj,Spergel:2006hy} as well as large scale structure\cite{Ahn:2012fh}, the building of concrete models of inflation which are consistent with the known precepts of quantum field theory, {\em and} have some measure of naturalness has been a vexing problem since inflation was first posited. The main issue is how to keep quantum corrections from disturbing the required flatness of the potential $V(\phi)$ for the inflaton $\Phi$, as measured by the slow-roll parameters $\epsilon= M_{\rm Pl}^2\slash 2\ \left(V'(\Phi)\slash{V(\Phi)}\right)^2,\ \eta =  M_{\rm Pl}^2 \left(V''(\Phi)\slash{V(\Phi)}\right)$. One exception to this situation occurs when $\Phi$ is the pseudo-Nambu-Goldstone boson (PNGB) of a broken symmetry. In this case, there can be a residual shift symmetry that protects the flatness of the potential from obtaining quantum corrections. It is exactly this property of PNGB's that was exploited in Natural Inflation (NI)\cite{Freese:1990rb,Adams:1992bn,Freese:2014nla}. A PNGB $\Phi$ was taken to be the inflaton with a potential of the form 

\begin{equation}
\label{eq:axionpotential}
V(\Phi) = \Lambda^4 \left(1+\cos\frac{\Phi}{f}\right),
\end{equation}
where $\Lambda$ and $f$ are mass scales that can be fixed by matching to observations.

If we now ask that we have at least 65 e-folds of inflation and that for modes leaving the horizon during the last 55-60 e-folds we arrive at values of the spectral index $n_s$ and the ratio of scalar to tensor amplitudes $r$ consistent with Planck\cite{Ade:2015lrj} and BAO\cite{Blake:2011en} data then $f>M_{\rm Pl}$ is required. This is problematic on a number of levels. First and foremost, given that the field $\Phi$ moves a distance greater than $f$ in field space during inflation, we need to cope with field values greater than the Planck scale. Generically, this would imply that Planck suppressed higher dimension operators could not be neglected or, to put it another way, the effective field theory treatment of $\Phi$ would break down. A second issue is that while axions exist in abundance in string theory, super-Planckian axion decay constants are difficult to obtain\cite{Svrcek:2006yi} within this framework. In all, it is safe to say that from both a theoretical and observational point of view, the requirement that $f>M_{\rm Pl}$ gives rise to concerns about just how ``natural'' Natural Inflation can be.

Ideas such as N-flation\cite{Dimopoulos:2005ac}, aligned inflation\cite{Kim:2004rp}, which use the plethora of axions found in string theory to either reduce the effective value of $f$ by the number $N$ of axions or balance the different decay constants of two axions to obtain a super-Planckian {\em effective} decay constant have been suggested as potential ways out of this dilemma (see also Refs.\cite{Burgess:2014oma,Kaloper:2011jz,delaFuente:2014aca,Croon:2014dma,Czerny:2014wza,Czerny:2014xja,Czerny:2014qqa} and, in the context of multi-field inflation, Refs.\cite{Liddle:1998jc,Copeland:1999cs}). Other types of string constructions give rise to the axion monodromy\cite{Silverstein:2008sg} scenarios in which the field manifold is such that a sub-Planckian field excursion can give rise to an effective super-Planckian one. 
 
In all of these constructions, the hope is that by dint of either the field content or the vagaries of the field manifold, the need for a fundamental super-Planckian axion decay constant can be obviated. In this work we offer a different approach, one that uses the non-equilibrium dynamics of PNGB's to show that consistent inflation can be achieved in NI even with $f<M_{\rm Pl}$. The main observation leading us to this result is that the potential in eq.(\ref{eq:axionpotential}) has regions where the frequency $\omega_k$ for modes of comoving wavenumber $k$ is imaginary; these give rise to {\em spinodal} instabilities\cite{Cormier:1999ia}. The equation of motion for these modes is given by
 
\begin{equation}
\label{eq:spinodalmodeqns}
\left[\frac{d^2}{dt^2} + 3 H(t)\frac{d}{dt} + \left(\frac{k^2}{a(t)^2}- \frac{\Lambda^4}{f^2}\cos\left(\frac{\Phi}{f}\right)\right)\right]g_k(t)=0,
\end{equation}
where $a(t)$ is the scale factor of the FRW geometry as usual, $H(t)$ the corresponding Hubble parameter and we have decomposed the field $\Phi$ as

\begin{equation}
\label{eq:modedecomp}
\Phi(\vec{x},t) = \phi(t) + \psi(\vec{x},t),\ \psi(\vec{x},t) = \frac{1}{\sqrt{V}}\sum_{\vec{k}} g_k(t) e^{-i\vec{k}\cdot\vec{x}}.
\end{equation}
Thus $\phi(t)$ is the zero momentum mode and we have used a box of comoving volume $V$ for our momentum expansion. We see from eq.(\ref{eq:spinodalmodeqns}) that for $0<\phi\slash f <\pi\slash 2$ and for low enough values of $k\slash a(t)$, the mode $g_k(t)$ is unstable. An inflationary period ensures that more and more modes will be redshifted into the instability region, though the dominant effect on the background evolution comes from the most unstable modes, i.e., the ones that were near the unstable regime early on during inflation. 

The main way in which the spinodal instability makes itself felt is through the growth of the two-point function of the fluctuations $\langle \psi(\vec{x},t)^2\rangle$:

\begin{equation}
\label{eq:twopoint}
\langle \psi(\vec{x},t)^2\rangle = \int \frac{d^3 k}{\left(2\pi\right)^3}\ \left | g_k(t)\right |^2.
\end{equation}

In fact due to the continuing influx of unstable modes during inflation, the two-point function will become non-perturbatively large\cite{Cormier:1999ia}. How can we tame this growth so as to be able to understand how the fluctuations influence the evolution of the zero mode? One way to do this is via the so-called Hartree approximation, wherein interactions such as $\psi^{2 n}$ are approximated by $\psi^{2 n}\rightarrow a_n \langle \psi^2\rangle^n +b_n \langle \psi^2\rangle^{n-1} \psi^2$ and $\psi^{2 n+1}\rightarrow c_n \langle \psi^2\rangle^n \psi$ with the coefficients $a_n, b_n, c_n$ found by demanding that the approximate theory reproduce the perturbative results for the zero, one and two-point functions of the full theory. This has the effect of rendering the interacting theory a Gaussian one, but with remnants of the interaction still present in the self-consistent computation of $\langle \psi^2\rangle$. Diagrammatically, the Hartree approximation corresponds to summing the ``cactus'' diagrams of the theory. 

Turning back to the case of Natural Inflation, the Hartree approximation involves inserting eq.(\ref{eq:modedecomp}) into eq.(\ref{eq:axionpotential}), expanding the cosines and sines, and then making the
following replacements\cite{Cormier:1999ia}: 
\begin{eqnarray}
\label{eq:hartreereplacements}
&& \cos \left(\frac{\psi }{f}\right)\rightarrow \left( 1-\frac{\left( \psi
^{2}-\left\langle \psi ^{2}\right\rangle \right) }{2f^{2}}\right) \exp\left( -%
\frac{\left\langle \psi ^{2}\right\rangle }{2f^{2}}\right),\nonumber\\
&& \sin \left(\frac{%
\psi }{f}\right)\rightarrow \frac{\psi }{f}\exp\left( -\frac{\left\langle \psi
^{2}\right\rangle }{2f^{2}}\right).  
\end{eqnarray}
The equations for the field $\phi ,$ and the fluctuation modes $g_{k}$
coupled to the scale factor $a\left( t\right) $ are 
\begin{eqnarray}
&&\ddot{\phi}+3H(t)\dot{\phi}-\frac{\Lambda^{4}}{f}\exp\left(-\frac{%
\left\langle \psi ^{2}\right\rangle }{2f^{2}}\right)\sin\left(\frac{\phi}{f}\right)=0, 
\label{eq:eqofmotion} \nonumber \\
&&\ddot{g}_{k}+3H(t)\dot{g}_{k}+ \\ 
&&\left[\frac{k^{2}}{a^{2}(t)}-\frac{%
\Lambda^{4}}{f^{2}}\exp\left(-\frac{\left\langle \psi ^{2}\right\rangle }{%
2f^{2}}\right)\cos\left(\frac{\phi }{f}\right)\right] g_{k}=0.  \nonumber
\end{eqnarray}

The effective Friedmann equation for the scale factor is obtained by use of semiclassical gravity, i.e. by using $\left\langle T_{\mu \nu }\right\rangle$ to source the Einstein equations (here $ \tilde{M}_{\rm Pl}= M_{\rm Pl}\slash \sqrt{8\pi}$ is the reduced Planck mass): 

\begin{eqnarray}
&& H^2(t)=\frac{1}{3\tilde{M}_{\rm Pl}^{2}}\left[\frac{1%
}{2}\dot{\phi}^{2}+\frac{1}{2}\langle \dot{\psi}^{2}\rangle +\frac{1}{2a^{2}}%
\left\langle (\vec{\nabla}\psi )^{2}\right\rangle \right . \nonumber \\
&& \left .+\Lambda^{4}\left( 1+\cos\left(\frac{\phi}
{f}\right)\exp\left(-\frac{\left\langle \psi ^{2}\right\rangle }{2f^{2}}\right)\right) %
\right] , 
\end{eqnarray}

How do the inflationary dynamics of this system differ from the usual dynamics of natural inflation (what we will henceforth call ``vanilla'' NI)? We plot the evolution of the zero mode, the fluctuations $\langle \psi^2 \rangle$ and the Hubble parameter below as functions of $\tau \equiv \mu t$, where we define $\mu = \Lambda^2\slash f,\ \lambda = \Lambda\slash f,\ \alpha = f\slash \tilde{M}_{\rm Pl}$. The figures below use $\alpha= 0.5,\ \lambda=10^{-4},\ \phi(0)\slash f=2.85\times 10^{-10}$, parameters for which vanilla NI would not give enough e-folds to give a viable inflationary scenario. 

\begin{figure}[ht]
\centering
\subfloat[]{%
\includegraphics[scale=0.54]{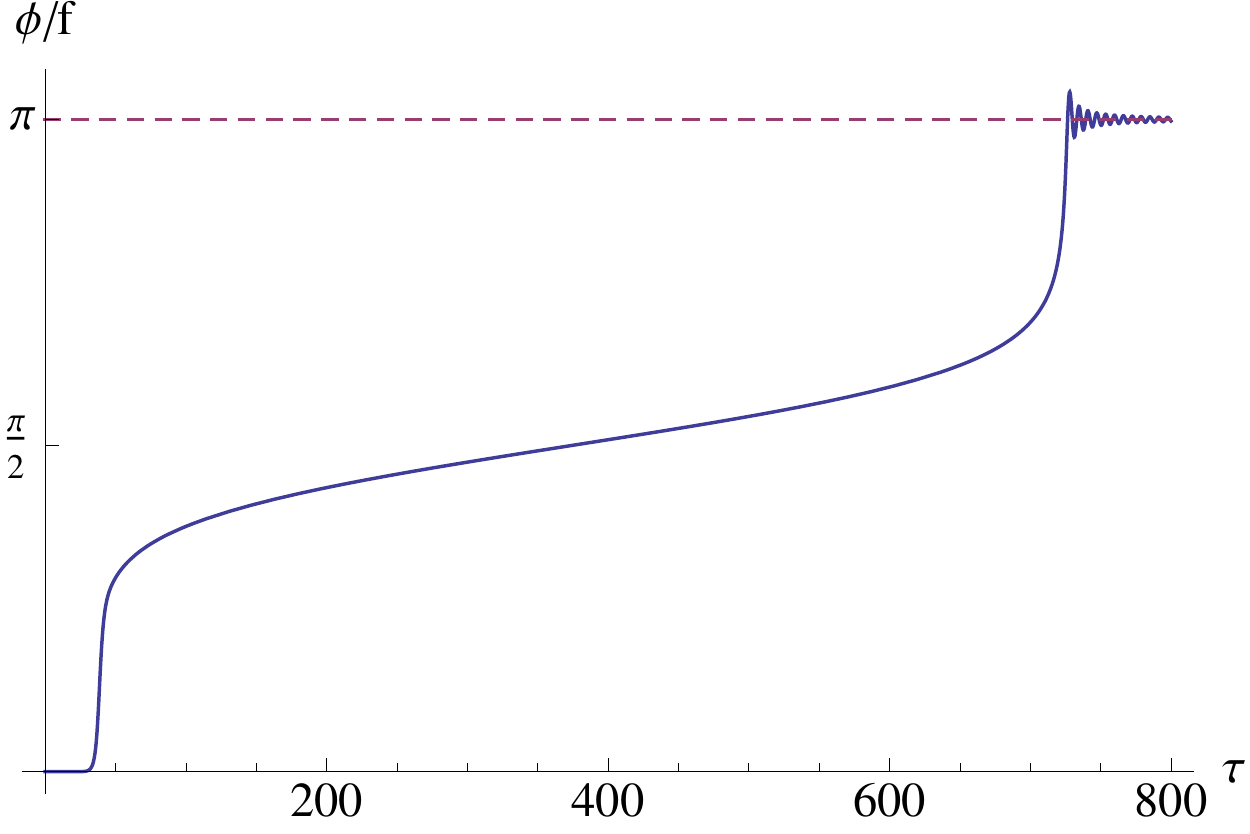} 
\label{fig:phioftau}}
\quad
\subfloat[]{%
\includegraphics[scale=0.54]{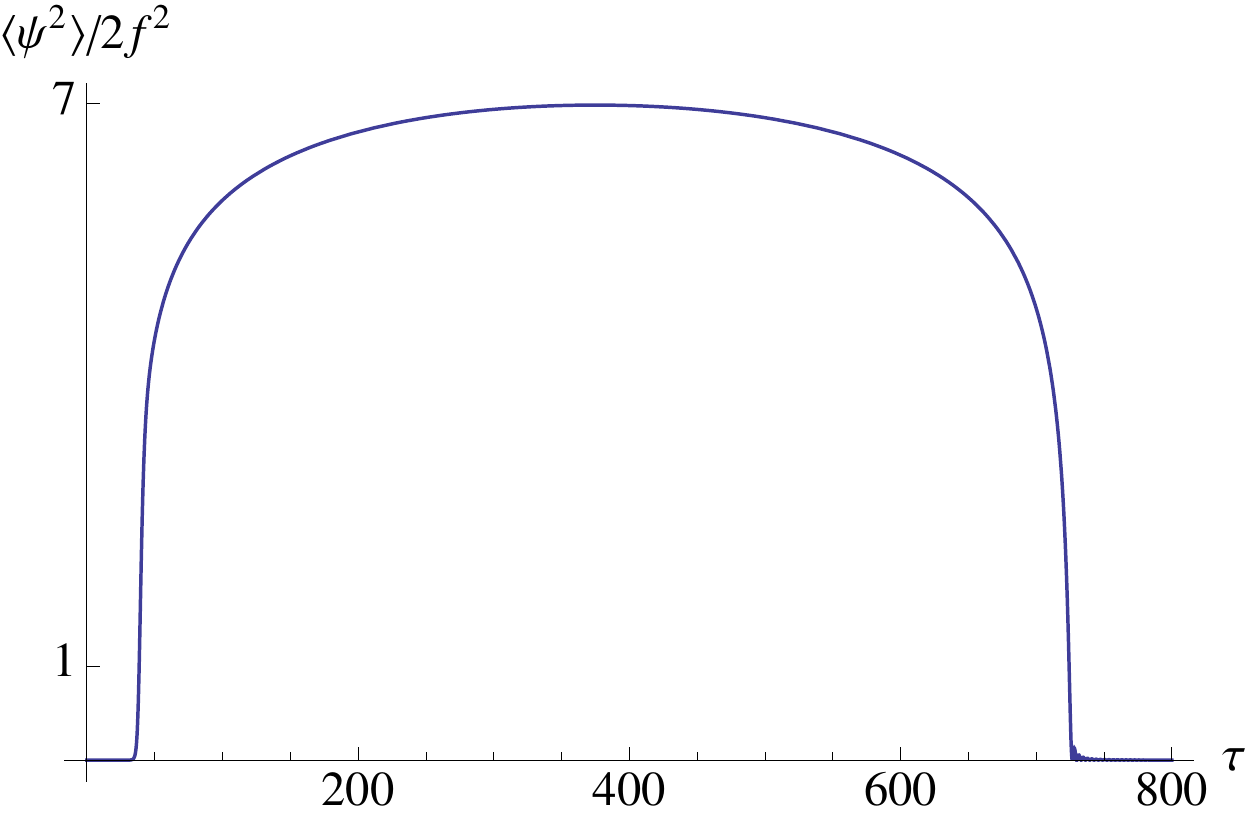}
\label{fig:fluctsoftau}}
\quad
\subfloat[]{%
\includegraphics[scale=0.54]{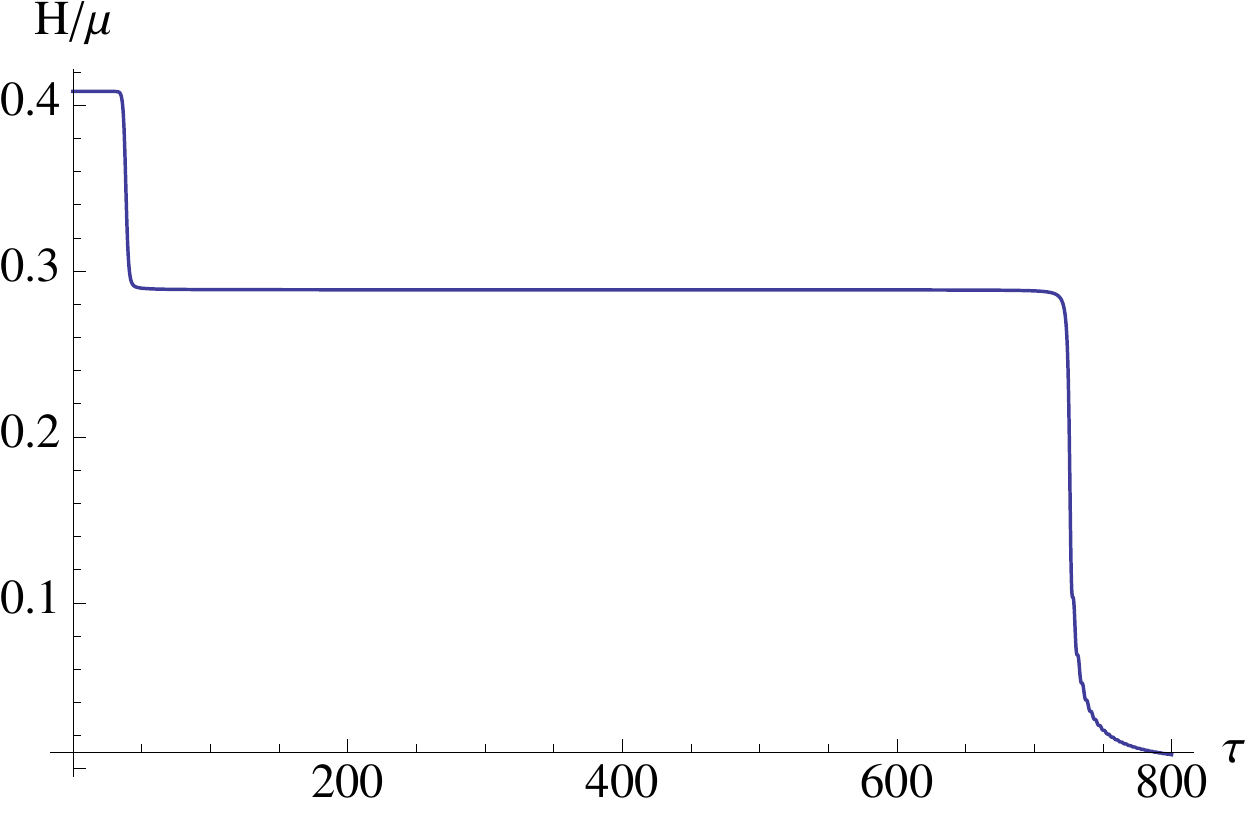}
\label{fig:hoftau}}
\caption{(a) The field $\phi$ in units of $f$, (b) the fluctuations $\langle \psi^2 \rangle\slash 2 f^2$ in units of $\lambda^{-4}$, (c)  The Hubble parameter $H$ in units of $\mu =\Lambda^2\slash f$ as functions of $\tau = \mu t$.}
\label{fig:rowfigs}
\end{figure}

\begin{figure}[!h]
\centering
\includegraphics[scale=0.53]{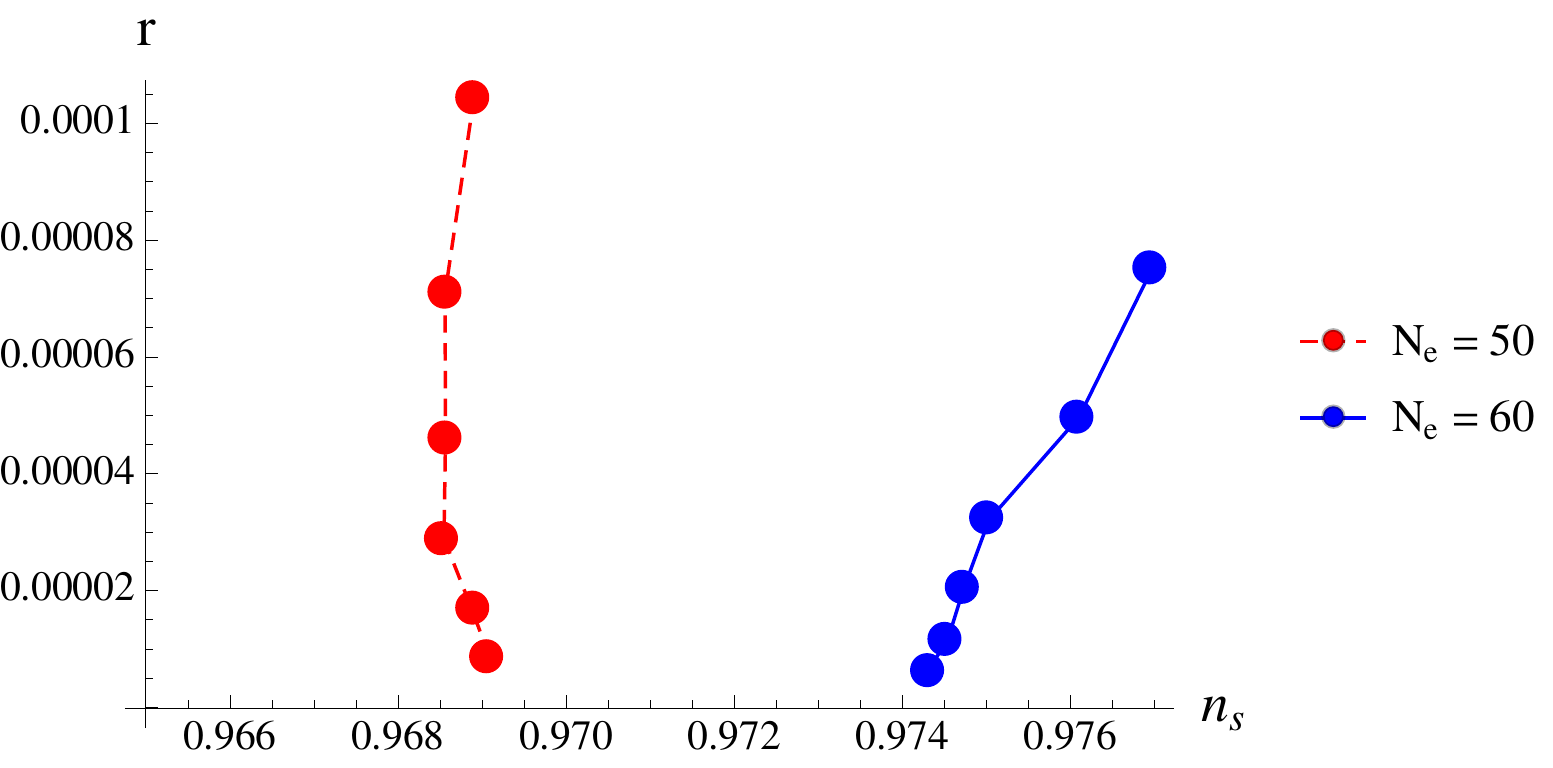}
\caption{The ratio of tensor to scalar fluctuations $r$ as a function of the spectral index $n_s$, for models with  total number of e-folds $\sim 200$. $N_{\rm e-folds}$ in the figure corresponds to the number of e-folds before inflation ends. Higher values of $\alpha = f\slash \tilde{M}_{\rm Pl}$ correspond to the points with larger values of $r$.}
\label{fig:nsralpha}
\end{figure}


As can be seen from figure \ref{fig:rowfigs}, the zero mode $\phi(t)$ starts off following vanilla dynamics. This continues until the unstable modes have made the fluctuations $\langle \psi^2 \rangle\slash 2 f^2\sim {\cal O}(\lambda^{-4})$, at which point the dynamics become dominated by the fluctuations and enter the spinodal regime. A {\em second} phase of inflation begins, dominated by a potential energy density given by the value of the potential at the spinodal point, which for NI is half the original potential energy density. The total number of e-folds for these parameter choices is slightly over $200$ and the last $60$ e-folds are wholly contained within the second phase of inflation. 

As shown in ref.\cite{Cormier:1999ia},the power spectrum can be computed from
\begin{eqnarray}
\label{eq:powerspectrum}
&& {\cal P}(k) = \frac{1}{300 \pi^2 \tilde{M}^2_{\rm Pl}} \frac{\langle V(\phi+\psi)\rangle \langle \partial_{\psi} V(\phi+\psi)\rangle}{\left(\dot{\phi}^2 +\langle \dot{\psi}^2 \rangle\right)^2}\nonumber\\
&& \langle \dot{\psi}^2 \rangle = \int \frac{d^3 k}{\left(2\pi\right)^3}\ \left | \dot{g}_k(t)\right |^2,
\end{eqnarray}
which is evaluated at the time $t_k$ at which the mode first crosses the horizon: $k=a(t) H(t)$. For our potential this becomes
\begin{equation}
\label{eq:ourpowerspectrum}
 {\cal P}(k) =\frac{\alpha^2 \lambda^4}{600 \pi^2}\
 \frac{\left[1-e^{-4\tilde{\sigma}^2}\cos\left(2\tilde{\phi}\right)\right] \left[1+e^{-\tilde{\sigma}^2}\cos\left(\tilde{\phi}\right)\right]}{\left(\dot{\phi}^2 +2 \dot{\tilde{\sigma}}^2\right)^2},
\end{equation}
where we have defined $\tilde{\phi} \equiv \phi\slash f$, $\tilde{\sigma}^2$ as $\langle\psi^2\rangle\slash 2 f^2$ and $\dot{\tilde{\sigma}}^2$ as $ \langle \dot{\psi}^2 \rangle\slash 2 f^2$.
We can use this to read off the scalar spectral index as well as the ratio of tensor to scalar fluctuations $r$; for the above parameter choices, $n_s = 0.9736$ and $r=6\times 10^{-6}$ if the pivot scale $k=0.002\ {\rm Mpc}^{-1}$ leaves $60$ e-folds before the end of inflation, while if this scale leaves at $50$ e-folds before the end of inflation, we have $n_s = 0.9689$ and $r=8.6\times 10^{-6}$. While the former value of $n_s$ is a little high compared to the central value of $0.9655 \pm 0.0062$ ($68\%$ CL) found by Planck\cite{Ade:2015lrj}, both are well within the $95\%$ CL region in the $n_s-r$ plane. In figure \ref{fig:nsralpha} we show how $n_s,\ r$ vary within a set of models with the same number $\sim 200$ total number of e-folds but different values of $\alpha = f\slash \tilde{M}_{\rm Pl}$ ranging from $0.5$ to $1$. The trends we see are the same as for the model considered above: $n_s$ is consistent with data and $r$ is unobservably small. In light of the issues appearing in the interpretation of the BICEP2\cite{Ade:2014xna} results, the smallness of $r$ does not overly concern us at this point\cite{Flauger:2014qra,Mortonson:2014bja,Ade:2015tva}. There may be models that result in a larger $r$, for $\alpha < 1$, while keeping an acceptable value of $n_s$.

What we have exhibited here is a choice of parameters for which all constraints on inflationary models are satisfied starting with the potential for NI, yet no super-Planckian field excursions occur; in fact $f\sim M_{\rm Pl}\slash 10$ ($ M_{\rm Pl}$ is the {\em actual} Planck mass). The details of whether the spinodal phase of inflation occurs, and for how long depends relatively sensitively on the initial conditions\cite{Cormier:1999ia} and the values for both $n_s$ and $r$ will reflect this, making it more difficult to give hard predictions for these values than in vanilla NI. 

How certain are we of the details of the spinodal dynamics? It is clear that {\em something} happens due to the growth of fluctuations while the zero mode evolves within the spinodal regime. The question is whether the Hartree approximation captures enough of this dynamics to be reliable. Unfortunately, the Hartree approximation is not a controlled one, so it is hard to quantify its accuracy. However, it is the $N\rightarrow 1$ limit of the large $N$ approximation which when applied to inflation\cite{Boyanovsky:1997xt} exhibits behavior remarkably similar to that seen here. The essential difference is that the presence of Goldstone modes in the large $N$ case leads to the spinodal line lying at the bottom of the potential well. But long wavelength modes grow large and can influence the dynamics in that situation, so we believe that the Hartree approximation is capturing the most essential aspects of the evolution of the system. 

Another potential issue with the Hartree approximation is that it does not take scattering into account and it might be argued that such effects might deplete the long wavelength modes by scattering them into shorter wavelength ones. While we cannot answer this conclusively there has been some work on non-spinodally unstable $\lambda \phi^4$\cite{Aarts:2000wi} that argues that the two-point functions we are following here behave in qualitatively similar ways when scattering effects are accounted for. 

Note that the modes that are driving the spinodal instabilities are the extremely super-horizon ones, those that would not yet have re-entered our horizon. Thus, the integrals defining the various expectation values used in our analysis are dominated by these modes while the modes that will appear in the CMB sky can be treated in the usual perturbative fashion.

We would argue then that it is premature to rule out Natural Inflation models with $f< M_{\rm Pl}$. When the quantum dynamics of the spinodally unstable modes is taken into account, a second phase of inflation ensues, driven by the energy density of these modes. This second phase alone can give rise to a sufficient number of e-folds to solve the horizon and flatness problem and, for some values of the parameters, $n_s$ and $r$ can take on values consistent with current constraints. A more systematic search of the parameter space of spinodal NI models needs to be done; this is currently being undertaken by us. But we are heartened to see that there are {\em some} parameters for which the spinodal phase gives rise to a consistent inflationary model.

Finally, we note that given the most recent PLANCK results\cite{Ade:2015lrj}, inflationary models where the inflaton starts in the concave part of the potential seem to be prefered. Given that the inflaton will evolve towards a minimum of this potential, a spinodal region will necessarily {\em always} exist in such models. Thus our analysis, rather than being a ``one-off'' for a particular model, will, in fact, need to be applied to {\em all} such models.

\begin{acknowledgments}
A.~A. and B.~R. were supported in part by DOE grant DE-FG03-91ER40674, while R.~H. was supported in part by the Department of Energy under grant DE-FG03-91-ER40682. The authors would like to thank Marius Millea and Andrew Scacco for useful discussions. R.~H also thanks the cosmology group at UC Davis for hospitality while this work was in progress. 
\end{acknowledgments}

\bibliography{SpNI}

\end{document}